\documentclass[twocolumn,prd,amsfonts,showpacs,epsfig,floats,superscriptaddress,floatfix,nofootinbib]{revtex4}
\usepackage{graphicx}
\usepackage{dcolumn}
\usepackage{bm}
\usepackage{longtable}
\usepackage{amsmath}
\usepackage{mathrsfs}
\usepackage{epstopdf}
\graphicspath{{./images/}}

\newcommand{\tbf}{\textbf}

\begin{document}
\bibliographystyle{apsrev}


\title{Late universe dynamics with scale-independent linear couplings in the dark sector}

\author{Claudia Quercellini}
\email[]{claudia.quercellini@uniroma2.it}
\affiliation{Dipartimento di Fisica, Universit\`a di Roma ``Tor Vergata'',
via della Ricerca Scientifica 1, 00133 Roma, Italy}
\author{Marco Bruni}
\email[]{marco.bruni@port.ac.uk} 
\affiliation{Institute of Cosmology and Gravitation, University
of Portsmouth, Mercantile House, Portsmouth PO1 2EG, Britain}
\affiliation{Dipartimento di Fisica, Universit\`a di Roma ``Tor Vergata'',
via della Ricerca Scientifica 1, 00133 Roma, Italy}
\author{Amedeo Balbi}
\email[]{amedeo.aalbi@roma2.infn.it}
\affiliation{Dipartimento di Fisica, Universit\`a di Roma ``Tor Vergata'',
via della Ricerca Scientifica 1, 00133 Roma, Italy}
\affiliation{INFN Sezione di Roma ``Tor Vergata'',
via della Ricerca Scientifica 1, 00133 Roma, Italy}
\author{Davide Pietrobon}
\email[]{davide.pietrobon@port.ac.uk}
\affiliation{Dipartimento di Fisica, Universit\`a di Roma ``Tor Vergata'',
via della Ricerca Scientifica 1, 00133 Roma, Italy}
\affiliation{Institute of Cosmology and Gravitation, University
of Portsmouth, Mercantile House, Portsmouth PO1 2EG, Britain}


\date{\today}

\begin{abstract}
We explore the dynamics of cosmological models with two coupled
dark components with energy densities $\rho_A$ and $\rho_B$  and constant equation of state (EoS) parameters $w_A$ and $w_B$. We
assume that the coupling is of the form $Q=H\, q(\rho_A,\rho_B)$,
so that the dynamics of the two components turns out to be  scale
independent, i.e.\ does not depend explicitly on the Hubble scalar
$H$. With this assumption, we focus on the general linear coupling
$q=q_o +q_A\,\rho_A+q_B\,\rho_B$, which may be seen as arising
from any $ q(\rho_A,\rho_B)$ at late time and leads in general to
an effective cosmological constant. In the second part of the
paper we consider observational constraints on the form of the coupling from SN Ia data, assuming that one of the components is cold dark matter (CDM), i.e.\ $w_B=0$, while for the other the EoS parameter can either have a standard ($w_A>-1$) or phantom ($w_A<-1$) value. We find that the constant part of the coupling function is unconstrained by SN Ia data and, among typical linear coupling functions, the one proportional to the dark  energy density $\rho_{A}$ is preferred in the strong coupling regime, $|q_{A}|>1$.  Models with phantom $w_A$  favor a positive coupling function, increasing $\rho_A$.  In  models with standard $w_A$, not only a negative coupling function is allowed, transferring energy to CDM, but  the uncoupled sub-case falls at the border of the likelihood.
\end{abstract}

\maketitle


\section{Introduction}
\label{intro}
The overall density of the observed universe, the growth of structures and their clustering properties cannot be explained by known forms of matter and energy \cite{2006Natur.440.1137S,Khalil:2002}. In addition, cosmic microwave background (CMB) anisotropy observations show that the total  density  is close to critical, so that the gap between known and unknown cannot be accounted for by curvature \cite{Spergel:astro-ph/0603449,Dunkley:2008}. Finally, 
observations of type Ia Supernovae (SNe) \cite{Riess:1998,Perlmutter:1999,Riess:astro-ph/0611572}, baryon acoustic oscillations (BAO) \cite{Eisenstein:astro-ph/0501171,Percival:2007p358}, and integrated Sachs-Wolfe (ISW) effect\cite{Pietrobon:2006,Giannantonio:2008}   tell us that the universe expansion is currently accelerating.
To explain these facts cosmologists  need to assume the existence of a dark sector in the theory \cite{Copeland:2006}, whose general properties have then to be tested against observations, i.e.\ with parameters that have to be  deduced from indirect evidences. One possibility is that the dark sector is accounted for, partly or in full, by a modified gravity theory\footnote{See e.g.\ \cite{Durrer:2008} and other articles in the same special issue on dark energy.}, while a more conventional approach is to assume that gravity is well described by general relativity, with the dark sector made up of an unusual energy momentum tensor.

 In the currently prevailing scenario, the dark sector consists of two distinct contributions. One component, {\em cold dark matter} (CDM), accounts for about one third of the critical density
\cite{Percival:2007p1293} and is needed to explain the growth of inhomogeneities that we observe up to very large scales, as well as a host of other cosmological observations which goes from galactic scales, to clusters of galaxies, to redshift surveys. The other contribution, dubbed {\em dark energy}, accounts for the remaining two thirds of the critical density, and is needed to explain the observed late time acceleration of the universe expansion \cite{Perlmutter:1999,Riess:astro-ph/0611572}. CDM can be modeled as a pressureless perfect fluid, representing unknown heavy particles, collisionless and cold, i.e.\ with negligible velocity dispersion. 
In its simplest form, dark energy consists  of  vacuum energy density, i.e.\ a cosmological constant $\Lambda$. Taken together, $\Lambda$  and CDM make up for the the so-called concordance $\Lambda$CDM model \cite{Spergel:2003,Tegmark:2004}. This simple model fits observations reasonably well, but lacks a sound explanation in terms of fundamental physics, and a number of alternatives have been proposed. 
In general, dark energy can  be modeled as a perfect fluid with an equation of state (EoS from now on) that violates the strong energy condition \cite{Visser:1997aa}, such that it can dominate at late times and  have sufficiently negative pressure to account for  the observed accelerated expansion. Scalar fields can also be formally represented as perfect fluids (see e.g.\ \cite{Bruni:1992p718} and refs.\ therein).
In a more exotic version, dubbed {\it phantom energy} \cite{Caldwell:2002p1819,Caldwell:2003p1814}, the EoS also violates the null energy condition \cite{Visser:1997aa},  leading to the growth in time of the energy density with the cosmic expansion. Finally, another rather radical alternative to $\Lambda$CDM is to assume a single unified dark matter (UDM), able to mimic the essential features of $\Lambda$CDM that are necessary to build a viable cosmology. For example, in \cite{Balbi:astro-ph/0702423} we have considered observational constraints on a  UDM model with an ``affine" EoS, i.e.\ such that the pressure satisfies the affine relation  $P=P_o+\alpha\rho$  with the energy density \cite{Ananda:astro-ph/0512224,Ananda:2006}. This model is a one parameter ($\alpha$) generalization of  $\Lambda$CDM, with the latter recovered for $\alpha=0$. There is no need to assume {\it a-priori} a $\Lambda$ term in Einstein equations, because the EoS $P=P_o+\alpha\rho$  leads to an effective cosmological constant with $\Omega_\Lambda =-8\pi G P_o/[3 H^2_o (1+\alpha)]$.
The problem is thus shifted from justifying a $\Lambda$ term in Einstein equations to that of justifying the assumed EoS:
a possible justification of this affine model can be given in terms of scalar fields, either of quintessence or k-essence type \cite{Quercellini:2007}. This type of model escapes typical constrains on many UDM models \cite{Sandvik:2004p316} (but cf.\ e.g.\ \cite{Gorini:2007}) because, for a given homogeneous isotropic background expansion, it allows multiple phenomenological choices for the speed of sound of the perturbations \cite{Pietrobon:2008}. 

 In models of the dark sector consisting of two components, dark matter and dark energy  are usually assumed to interact only through gravity, but they might exhibit other interactions without violating observational constraints \cite{Kunz:2007p2562}.
Exploiting this degeneracy, here we depart from the standard scenario, and assume a cosmological model where the dark sector is made up of two coupled dark components, each described as a perfect fluid with its own constant EoS parameter $w$.  
This choice allows for the possibility that the observed evolution of the universe, although reasonably well explained by the $\Lambda$CDM model, is actually due to the dynamics of two rather general coupled components, possibly alleviating the so-called ``coincidence problem", $\Omega_\Lambda \approx \Omega_{CDM}$, typical of the standard model \cite{Copeland:2006}.


In this paper, our first aim is to characterise  the dynamics of our cosmological model with  the two general coupled components, taking into account general forms of interaction, parameterized in terms of a late time function $Q$ linear in the energy densities, Eqs.\ (\ref{Coup}-\ref{coupling}). To this end we will use standard dynamical system techniques \cite{KArrowsmith:1992,Wainwright:1997}, which are now rather common in the analysis of cosmological models, see e.g.\ \cite{Wands:1993,Bruni:1993,Amendola:1993,Bruni:1994,Bruni:1995a,Bruni:1995b} and \cite{Copeland:1998p1287,Ananda:astro-ph/0512224,Ananda:2006,Bohemer:2008}. To our knowledge, such an exhaustive analysis has not been carried out yet, although several sub-cases have been considered \cite{Majerotto:2004,Olivares:2006,Guo:2007,Bohemer:2008,Quartin:2008, pettorino:2008,Barrow:2006}. In our study, we restrict ourselves to the evolution of a homogeneous, isotropic cosmological background, leaving aside the question of what the effects of coupling could be in anisotropic models \cite{Ananda:2006}, or when general perturbations are present \cite{Valiviita:2008p2415,Dunsby:1992p713}. It is however worth noticing that, thanks to the particular form of coupling we choose, our analysis of  the dynamics of the two components is valid in any theory of gravity, because is based only on the conservation equations, and not on specific field equations. 

Secondly, as a way to gain some physical insight on the likelihood of some specific coupling models, we also explore the constraints on the predicted luminosity distance modulus derived from type Ia Supernovae observations, using a Monte Carlo Markov Chain (MCMC) approach. Needless to say, this is not intended as a full-fledged cosmological parameter estimation for these models, but only as a first exploration of the parameter space to rule out those models which are manifestly in contrast with observations. This analysis requires the use of the Friedmann equation, hence general relativity is assumed as the valid theory of gravity.

The paper is organized as follows. In Section II we study the general dynamics and solve the equations for the evolution of the coupled dark components; in Section III we focus on the cosmological  effects of changing the parameters of the coupling term; in Section IV we derive constraints from the observation of type Ia SNe on some specific sub-classes of models; finally, in Section V, we present the main conclusions of our work.
 
\section{dynamics of dark components}
\label{coupledfluids}
\subsection{Linear scale-free coupling}
In general relativity, assuming a a flat Robertson-Walker universe, the dynamics is subject to the Friedmann constraint
\begin{equation}
\label{einstein1}
H^2=\frac{8\pi G}{3}\,\rho_T\,,
\end{equation}
where $\rho_T$ is the total energy density of the various components. Beside  baryons and radiation, $\rho_T$ includes  any other component contributing to  the dark sector, i.e.\  that part of the total energy-momentum tensor that in the context of general relativity is needed to explain the observed universe, in particular the CMB  \cite{Spergel:astro-ph/0603449,Dunkley:2008}, structure formation  \cite{2006Natur.440.1137S,Khalil:2002} and the late time acceleration of the expansion \cite{Riess:1998,Perlmutter:1999,Riess:astro-ph/0611572,Eisenstein:astro-ph/0501171,Percival:2007p358,Pietrobon:2006,Giannantonio:2008}. The dynamics itself is described by the evolution of the Hubble expansion scalar $H=\dot{a}/a$, given by the Raychaudhuri equation
\begin{equation}
\label{ray}
\dot{H}=-H^2 -\frac{4\pi G}{3} (1+3\, w_T)\rho_T.
\end{equation}
This is coupled to the evolution equations for the energy density  of each of the matter components contributing to $\rho_T$. Since $\dot{H}+H^2=\ddot{a}/a$, with $a(t)$ the usual metric scale-factor (which we assume normalized to its present value), acceleration is achieved whenever $w_T=P_T/\rho_T<-1/3$, as it is well  known. 

The  standard $\Lambda$CDM model assumes two dark components: the pressureless cold dark matter (CDM), with $w_{DM}=0$, and the cosmological constant $\Lambda$ with $w_\Lambda=-1$. CDM  is needed to fill the gap between the baryon abundance and the amount of matter that is needed to explain the rotation curve of galaxies and structure formation in general, as well as to allow for a  vanishing curvature model. In the context of general relativity, and under the Robertson-Walker homogeneus and isotropic assumption (see e.g.\ \cite{Celerier:2007} for  alternatives), a cosmological constant $\Lambda$ is the simplest possible form of dark energy (DE) needed to generate  the observed low redshift acceleration. While this simple scenario is preferred from the point of view of model comparison and selection \cite{Balbi:astro-ph/0702423},  because of  the low number of parameters, from a theoretical perspective is oversimplified, and it is worth exploring alternatives, even if purely phenomenological. 

Here we shall consider 
  two general  coupled dark components with energy densities  $\rho_{A}$ and $\rho_{B}$.  Since we want to introduce a rather general type of coupling, focusing our analysis on its effects,  we shall assume the simplest possible form for the EoS of these two dark components, i.e.\ we will assume that the EoS parameters $w_{A}$ and $w_{B}$ are constant. On the other hand, we shall not {\it a priori} restrict our study to the sub-class of models where  one of the two components represents CDM with, for instance, $w_B=0$.  
  
  Due to the presence of the coupling,  the two  dark components  satisfy the balance equations
  \begin{eqnarray}
\label{cons}
\dot{\rho_{A}}+3H(1+w_{A})\rho_{A}&=&Q\\
\label{cons2}
\dot{\rho_{B}}+3H(1+w_{B})\rho_{B}&=&-Q\,.
\end{eqnarray}
Even assuming the linear form for the  coupling $Q$ given in  (\ref{Coup}) and (\ref{coupling}) below, this model allows us to explore a large number of alternatives. Here we will focus on models for the homogeneous and isotropic background expansion, assuming that for those models that will fit current observational data it might always be possible to construct an appropriate perturbative scheme allowing for structure formation, for instance by assuming a vanishing effective speed of sound in one component. 

 The coupled dark components $\rho_A$ and $\rho_B$   could be in principle be taken to represent DE only, i.e.\ they could be two extra dark components contributing to $\rho_T$ in (\ref{einstein1}), {\it  in addition } to CDM. Leaving aside this possibility, and ignoring baryons and radiation as we will do in this Section, 
the sum of Eqs. (\ref{cons}-\ref{cons2}) gives the conservation equation for $\rho_T=\rho_{B}+\rho_{A}$.  A positive coupling term $Q$ corresponds to a transfer of energy from $\rho_{B}$ to $\rho_{A}$, and vice versa, but in general $Q$ doesn't need to have a definite sign. 

An interaction term between two components has been considered several times in literature, starting from Wetterich \cite{Wetterich:1998,Wetterich:1995} and Wands et al. \cite{Wands:1993,Copeland:1998p1287} in scalar field models, and has been analysed by Amendola in dark energy models \cite{Amendola:2000,Amendola:2001,Amendola:2003,Amendola:2004}, and for example recently in \cite{Majerotto:2004,Olivares:2006,Guo:2007,Bohemer:2008,Quartin:2008, pettorino:2008,Delamacorra:2008,Manera:2007}.

The coupling term $Q$ can take  any possible form $Q=Q(H,\rho_A,\rho_B,t)$. Here we shall consider the case of an autonomous ($t$ independent) coupling with a factorized $H$ dependence
\begin{equation}
\label{Coup}
Q=\frac{3}{2} H q(\rho_A,\rho_B).
\end{equation}
As we shall see below, with this assumption the effects of the coupling on the dynamics of $\rho_A$ and $\rho_B$ become effectively independent from the evolution of the Hubble scale $H$. For this reason, we may call this a  ``scale-independent" coupling.  Furthermore, with the decoupling of the dynamics of the two dark  components  from that of $H$,  the analysis of the next section is valid in any theory of gravity, because it is based on the conservation equations only: we don't need to use (\ref{einstein1})-(\ref{ray}), i.e. the field equations of general relativity.
Finally, we note that  any coupling of this type can be approximated at late times by a linear expansion:
\begin{equation}
\label{coupling}
q=q_{0}+q_{A}\rho_{A}+q_{B}\rho_{B}\;,
\end{equation}
where $q_A, q_B$ are dimensionless coupling constants, and $q_0$ is a constant coupling term with dimensions of an energy density\footnote{Strictly speaking, an expansion about today would lead to $q=\hat{q}_{0}+\hat{q}_{A}(\rho_{A}- \rho_{A0})+\hat{q}_{B}(\rho_{B} -\rho_{B0})$, but  constants  can always be re-defined in order to put the coupling $q$ in the form (\ref{coupling}). }.
In the following we shall analyse the dynamics arising from this general linear scale-independent coupling. Obvious sub-cases are: $q\propto \rho_T$ ($q_0=0, q_A=q_B$); $q\propto\rho_A$ ($q_0=0, q_B=0$); etc. We will come back to this in more detail in the next Section. 
Linear couplings have been frequently analysed in literature (\cite{Wetterich:1995,Amendola:2000,Majerotto:2004,Guo:2007},\cite{Multa:2007,Mainini:2007,Bohemer:2008,Quartin:2008}) both for mathematical simplicity, because they retain the linearity of system (\ref{cons}-\ref{cons2}) with no coupling, and because they can arise from string theory or Brans-Dicke-like Lagrangians after a conformal transformation of the metric.

\subsection{Analysis of the scale-free linear dynamics}
\label{stab}
\subsubsection{The linear dynamical system}
In order to proceed with the analysis of the dynamics of the dark components, let us change variables,  using  the total density $\rho_{T}=\rho_{B}+\rho_{A}$ and the difference $\Delta=\rho_{B}-\rho_{A}$. We also set 
\begin{eqnarray}
w_{+}=(w_{B}+w_{A})/2\,, & ~~ & w_{-}=(w_{B}-w_{A})/2\,, \\
q_{+}=(q_{B}+q_{A})/2\,, & ~~ & q_{-}=(q_{B}-q_{A})/2\,.
\end{eqnarray}
One reason for this choice is that ultimately the evolution of $\rho_T$  is the one that governs the general expansion law through (\ref{einstein1}) and (\ref{ray}).
In addition, thanks to the particular form of the coupling (\ref{Coup}) and assuming $H>0$, the dynamics can be made explicitly scale-independent, eliminating $H$ by adopting $N=\ln{(a)}$, the {\it e}-folding, as the independent variable.
Then, denoting with a prime the derivative with respect to $N$, the system  (\ref{cons}-\ref{cons2}) is transformed into 
\begin{eqnarray}
\label{constot}
\rho_{T}'+3\rho_{T}(1+w_{+})+3w_{-}\Delta&=&0 \qquad\qquad\qquad\qquad\qquad\quad\\
\label{cons2tot}
\Delta'+3\Delta(1+w_{+})+3w_{-}\rho_{T}&=&-3(q_{+}\rho_{T}+q_{-}\Delta+q_{0}).
\end{eqnarray}
 An effective EoS parameter $w_{eff}$ is implicitly defined from Eq.\ (\ref{constot}): when $w_-=0$ the two EoS coincide giving rise to a constant $w_{eff}=w_+=w_{A}$ and $\rho_T$ scales accordingly, as a standard barotropic perfect fluid, but in general 
 \begin{equation}
\label{weff}
w_{eff}=w_+ +w_-\frac{\Delta}{\rho_T}\
\end{equation}
changes with time. Notice  that we can also define, using (\ref{Coup}-\ref{coupling}) in (\ref{cons}), effective EoS parameters for the two components:
\begin{eqnarray}\label{wA}
w_{Aeff} & = & w_A -\frac{q_0+q_B\rho_B}{2\rho_A} -\frac{q_A}{2} \,,\\
w_{Beff} & = & w_B +\frac{q_0+q_A\rho_A}{2\rho_B} +\frac{q_B}{2} \,.
\label{wB}
\end{eqnarray}
From now on we will characterize the cosmological evolution of any of the energy densities as standard/phantom behaviour. As mentioned in the introduction, standard/phantom respectively correspond to an energy density which is either a decreasing or an increasing function of time (the scale factor or the {\it e}-folding {\it N}). The phantom behaviour arises in the presence of coupling from   an effective EoS parameter  $<-1$, which corresponds to the violation of the null energy condition \cite{Visser:1997aa} for that given energy density.  Thus, it follows from (\ref{weff}) and (\ref{wA}-\ref{wB})  that we can have a phantom behaviour in the total energy density $\rho_T$ as well as in one or both of the single components $\rho_A$ and $\rho_B$, and that in principle the effective EoS parameter of each of these can pass through the  $-1$ value, from phantom to standard or  vice versa. On the other hand, we will also refer to constant parameters such as $w_A$ and $w_B$ as having a standard/phantom value, respectively $w_A>-1$ or $w_A<-1$, because the corresponding fluid would evolve in that way in the case of no coupling.

 We will also refer to an ``affine'' evolution. As said  in the introduction, for an uncoupled component with energy density $\rho$ this arises from an affine EoS of the form $P=P_o +\alpha\rho$. Inserted in the   energy conservation equation this leads to  \begin{equation}
\label{affine}
 \rho=\rho_{\Lambda}+\rho_{0M}a^{-3(1+\alpha)}\, .
\end{equation}
Therefore, starting from the Friedmann equations (\ref{einstein1}-\ref{ray}) with no cosmological constant term, the affine EoS and energy conservation lead to  an effective cosmological constant $\rho_{\Lambda}$  plus an effective matter-like component  with constant EoS parameter $\alpha$ (a barotropic perfect fluid) and today's density $\rho_{0M}$ (cf.\  \cite{Ananda:astro-ph/0512224,Ananda:2006,Balbi:astro-ph/0702423,Quercellini:2007} for a detailed analysis of the cosmological dynamics arising in this case). As we will see, it turns out that there are  solutions of the system (\ref{constot}-\ref{cons2tot}) that evolve  according to (\ref{affine}).
 
In order to proceed with the analysis of Eqs.\ (\ref{constot}-\ref{cons2tot}) using standard dynamical system techniques \cite{KArrowsmith:1992}, it is convenient to write it as
\begin{equation}
\label{Xprime}
\bf{X}^\prime= { \bf J}\,  {\bf X} +{\bf C}\,,
\end{equation}
where the phase-space state vector $\bf X$ and the constant $\bf C$ are
\begin{equation}
\label{ defs}
{\bf X}=\left(\begin{array}{c}
      \rho_T    \\
      \Delta
\end{array}\right)\,, ~~~{\bf C}=\left(\begin{array}{c}
      0  \\
      -3q_0   
\end{array}\right)\,,
\end{equation}
and the matrix of coefficients ${\bf J}$ is given by
\begin{equation}
\label{J}
{\bf J}=\left(\begin{array}{rr}
      -3(1+w_+) & -3w_-   \\
      -3(w_-+q_+) & -3(1+w_++q_-) 
      \end{array}\right)  \;.
\end{equation}
Fixed points, if they exist, are solutions ${\bf X}_*$ of the equation $ { \bf J}\,  {\bf X}_* +{\bf C}={\bf 0}$ and,
given that the system (\ref{Xprime})  is linear, $\bf J$ is also the Jacobian of the system at these fixed points. These fixed points correspond to constant values of $\rho_{T}$ and $\Delta$ and in turn of $\rho_{A}$ and $\rho_{B}$,  that is to the emergence of an effective cosmological constant (when $\rho_T\not =0$, see below). Every constant form of energy is indeed alike the cosmological constant $\Lambda$, and plays exactly the same cosmological role: when dominates the evolution of the background, it drives an exponentially accelerated expansion, with an effective EoS parameter close to $-1$. Therefore, in this Section we will focus on the analysis of these fixed points.

Notice that - unlike the case with no coupling - there is no {\it a priori}  guarantee from the equations above  that $\rho_A$ and/or $\rho_B$, as well as  $\rho_T$, will  always be non-negative. However, one has to keep in mind that  $\rho_T$ must be non-negative because of the Friedmann constraint (\ref{einstein1}). This means that if  $\rho_T$ is vanishing for some value of $N$ ($a$), then at that point the assumption $H>0$, on the basis of which Eq.\ (\ref{Xprime}) is derived, is violated, and the solutions of (\ref{Xprime}) no longer correspond to solutions of the original coupled system of Eqs.\ (\ref{ray}) and (\ref{cons}-\ref{cons2}).

\subsubsection{Fixed points and stability analysis}
Even if system (\ref{Xprime}) is linear,  many different possibilities arise from the fact that it depends on five parameters\footnote{Only four of them are independent: one can always rewrite Eqs.\ (\ref{constot}-\ref{cons2tot}) renormalising the parameters to any of them, but we don't want to assume that any of $w_+, w_-, q_+, q_-, q_0$ is non null. In particular, the value of $q_0$ is irrelevant to the existence of the fixed points according to the eigenvalues of $\bf J$, but physically its value is relevant, because it determines the effective cosmological constant (\ref{effL}).}. There are two main cases, which we are now going to unfold.

\noindent
{\bf Case 1:} $\det({\bf J})\not =0$. In this case ${\bf J}^{-1}$ exists and there is a unique fixed point ${\bf X}_*=-{\bf J}^{-1} {\bf C}$. This can either be the origin in phase space, i.e.\ $\rho_{T*}=\Delta_*=0$, when $q_0=0$ (${\bf C}={\bf 0}$), or else this fixed point represents an effective cosmological constant,  with  $\rho_{T*}=\rho_\Lambda$, $\Delta_*=\Delta_\Lambda$ respectively given by
\begin{eqnarray}\label{effL}
\rho_\Lambda& = & 9\ \frac{w_-\ q_0}{\det({\bf J})}\,,  \\
\Delta_\Lambda & = & -9\ \frac{(1+w_+)\ q_0}{\det({\bf J})}\,,
\label{effD}
\end{eqnarray}
where
\begin{equation}
\label{detJ}
\det({\bf J}) =9\left[(1+w_+) (1+w_+ +q_-) -w_-(w_- +q_+)\right],\qquad
\end{equation}
 with corresponding values for $\rho_{A\Lambda}$ and $\rho_{B\Lambda}$.

In order to analyse the stability properties of system (\ref{Xprime}) at this fixed point, we now consider the eigenvalues of  $\bf J$.
Given that  $\det({\bf J})\not =0$, there are two non-zero eigenvalues, given by
\begin{equation}
\label{eigenvals}
\lambda_{\pm} =  \frac{{\rm tr}( {\bf J}) }{2} \pm\sqrt{D}\,, 
\end{equation}
where
\begin{eqnarray}
\label{trJ}
{\rm tr}( {\bf J}) & = & -3\left[2(1+w_+) +q_-\right]\,, \\
\label{D}
D & = &\left(\frac{{\rm tr}( {\bf J}) }{2}\right)^2 -\det({\bf J})= \\
 &=&9\left[ \left( \frac{ q_{-} }{2} \right)^{2}+w_{-}(q_{+}+w_{-}) \right]\,,
\end{eqnarray}
 and the value of the discriminant $D$ determines the three possible Jordan canonical forms of $\bf J$.
These three possible cases, to be further clarified in the next subsection, correspond to $D>0$, $D=0$, $D<0$, and are summarized below, with their sub-cases.
 
\noindent
{\bf Case 1a}: real distinct  eigenvalues ($D>0$). There are three sub-cases:  {\it i)} $\lambda_- < \lambda_+ <0$,  the fixed point is a stable node and both dark components have a standard behaviour, with decreasing energy densities;  {\it ii)} $\lambda_+>\lambda_->0$,  the fixed point is an unstable node and both dark components have a phantom behaviour, with increasing energy densities; {\it iii)} $\lambda_+>0>\lambda_-$, which requires $\det({\bf J})<0$; the fixed point is a saddle and both dark components  have a first  standard phase with decreasing energy density followed by a phantom phase. The connection between the phantom behaviour and sign of the eigenvalues will be rendered evident in the next Section, see Eq.\ (\ref{rhotot}) and related comments.

\noindent
{\bf Case1b}:  real equal eigenvalues ($D=0$). In this case $\lambda_+=\lambda_-=\lambda_0$ and the fixed point is an improper node, either stable (both dark components are standard) or unstable (both dark components are phantom). In both cases the two dark components, as well as $\rho_T$ and $\Delta$, follow a sort of  affine evolution, with a modification term, see Eq.\ (\ref{improper}). 

\noindent
{\bf Case 1c}:  complex eigenvalues ($D<0$). There are three sub-cases: {\it i)} if ${\rm tr}({\bf J})=0$ the fixed point is  a centre; {\it ii)} if ${\rm tr}({\bf J})>0$ the fixed point is  an unstable spiral; {\it iii)} if ${\rm tr}({\bf J})<0$ the fixed point is  a stable spiral. This last case is the most interesting, with $\rho_T$ converging to an effective cosmological constant via a series of oscillations. Notice that this case has to be dealt with care, as in the past $\rho_T=0$ ($H=0$) at some point, and  the time reversal of system (\ref{constot}-\ref{cons2tot})  should be considered prior to that. \\


\noindent
{\bf Case 2:}  $\det({\bf J}) =0$. In this case if $q_0\not =0$ (${\bf C} \not = {\bf 0}$) the system of linear equations $ { \bf J}\,  {\bf X}_* +{\bf C}={\bf 0}$ is in general inconsistent and there are no fixed points. Alternatively,  if  $q_0 =0$ (${\bf C}  = {\bf 0}$)  there is an infinite number of fixed points, each of them representing a possible asymptotic state depending on the initial conditions. These infinite number of fixed points is represented by a straight line in phase space, corresponding to a conserved quantity for system (\ref{Xprime}).
It turns out that there are three possible combinations of the four parameters $w_+, w_-, q_+, q_-$ that give $\det({\bf J}) =0$ and they are shown in Table \ref{detJ0}, classified as {\bf Case 2a, 2b} and {\bf 2c}.
For {\bf Case 2b}, $\rho_{T*}=0$ always, and either also $\Delta_{*}=0$ and then both $\rho_{A*}=\rho_{B*}=0$, or  $\Delta_{*}\not =0$ and then $\rho_{A*}=-\rho_{B*}=\Delta_*/2$, so that one of the two  is negative. Being the energy densities either null or negative, we can conclude that {\bf Case 2b} corresponds to a non-physical situation. {\bf Case 2a} and {\bf 2c} are more interesting, and are summarized in Table \ref{detJ0FP}. In both cases, 
 each fixed point on the line represents an effective cosmological constant. In particular, {\bf Case 2a} corresponds to $w_{A}=w_{B}=-1$, i.e.\ two cosmological constant-like components whose energy densities scale in a different way because of the coupling.  In addition, it turns out that for this very peculiar case $q_{0}$ can be non-zero. 
 
Regarding the stability analysis, in {\bf Case  2} $\det({\bf J})=0$ implies that one of the eigenvalues is null, namely $\lambda_-=0$ if ${\rm tr}( {\bf J})>0$ (and  vice versa) (cfr.\ Eqs.\ (\ref{eigenvals}) and (\ref{D})). This implies  that the total energy density, as well as the single dark components and $\Delta$, follow the affine evolution (\ref{affine}); we will comment further on this below Eq.\ (\ref{rhotot}). 
In this case, the non zero eingenvalue $\lambda_{-/+}$ for each sub-case is given in Table \ref{detJ0}. Table \ref{detJ0FP} gives instead the values the fixed points and the conditions for positive effective cosmological  constants.

Each of these fixed points is characterized by $\rho_{\Lambda}\propto \Delta_{\Lambda}$, which is equivalent to $\rho_{B\Lambda}\propto \rho_{A\Lambda}$. Explicitly,  for {\bf Case 2a} the energy densities  are related by $\rho_{B\Lambda}=-\rho_{A\Lambda}q_{A}/q_{B}$, while for {\bf Case 2c} $\rho_{B\Lambda}=-\rho_{A\Lambda}(1+w_{A})/(1+w_{B})$.  Note that the same proportionality law holds for the fixed point of {\bf Case 1} (cf.\ Eq.~(\ref{effL}-\ref{effD})). This behaviour mimics that of  scaling solutions \cite{Wands:1993,Copeland:1998p1287}, whose phase space typically admits fixed points where the contributions of the two fluids to the total energy density are constant (we will come back on this in Sec.~\ref{critpoints}). It is  easy to verify that at all these fixed points the effective EoS parameter (\ref{weff}) has  the value $w_{eff}=-1$.

\begin{table*}[htdp]
\begin{center}
\begin{tabular}{|c|c|c|}
\hline
Case & parameters & $\lambda_{+/-}$\\
\hline
 2a & $w_-=0$\,, $w_+=-1$\,, $\forall  q_+$\,, $\forall q_-$  & $-3q_-$\\
\hline
 2b & $w_-=0$\,, $q_-=-(1+w_+)$\,, $\forall  w_+$\,, $\forall  q_+$ & $-3(1+w_+)$\\
\hline
2c & $q_+=\frac{(1+w_+)(1+w_+ + q_-)-w_-^2  }{w_-}$\,, $w_-\not =0$\,,  $\forall  w_+$\,, $\forall  q_-$
& $-3[2(1+w_+) + q_-]$\\
\hline
\end{tabular}
\caption{The three possible combinations of parameters giving $\det({\bf J})=0$, cf.\ Eq.\ (\ref{detJ}). When $\det({\bf J})=0$, one of the eigenvalues (\ref{eigenvals}) vanishes, giving a constant mode, and the other is given for each case in  the third column here.
Thus $\rho_T$ follows the affine evolutions (\ref{affine}), i.e.\ there is a constant mode and a power law (in the scale factor a) mode, like that of a barotropic fluid; $\Delta$, $\rho_A$ and $\rho_B$ also have a a constant mode and the same power law mode.
\label{detJ0}}
\end{center}
\end{table*}%

\begin{table*}[htdp]
\begin{center}
\begin{tabular}{|c|c|c|c|c|}
\hline
Case & FPs &  $\rho_{A\Lambda}>0$ & $\rho_{B\Lambda}>0$ &  $\rho_{A\Lambda}>0$\,, $\rho_{B\Lambda}>0$ \\
\hline 
2a & $\rho_{\Lambda}=-\frac{q_-}{q_+} \Delta_{\Lambda}$ & $q_+>-q_-$ & $q_+<q_-$ & $\Delta_\Lambda >0$ and $-1<q_+/q_-<0$\,, or $\Delta_\Lambda <0$ and $0<q_+/q_-<1$\\
\hline
2c & $\rho_{\Lambda}=-\frac{ \Delta_\Lambda}{R}$ & $R>-1$ & $ R<1$ & $\Delta_\Lambda> 0$ and  $-1 <R<0$\,, or $\Delta_\Lambda< 0$ and  $0 <R<1$\\
\hline 
\end{tabular}
\caption{For {\bf Cases 2a} and {\bf 2c}, the fixed points and the conditions for positive effective cosmological constants, separated and combined, in both cases subject to the condition $\rho_{\Lambda}>0$.  For {\bf Case 2a} we have assumed $q_{0}=0$, although this is not a necessary condition. For {\bf Case 2c} we have defined $R=(1+w_+)/w_-$.\label{detJ0FP}}
\end{center}
\end{table*}%

\subsection{An equivalent equation and its solutions}

While the classification of the various possible phase portraits of system  (\ref{constot}-\ref{cons2tot})  is best given by  the representation (\ref{Xprime}) used in the previous Section, given that this is a linear system it is useful to consider the equivalent second order linear equation with constant coefficients, and its solutions. This helps interpreting the formalism of Sec.~\ref{stab} from a cosmological point of view. For this reason, we shall focus on the cases of main interest. 

First, notice that if $w_-=0$ then (\ref{constot}) decouples and implies 
\begin{equation}
\label{w-0}
\rho_T = \rho_c a^{-3(1+w_+)}\ , 
\end{equation}
with $\rho_c$ an integration constant, so that the total energy density can  either be standard or phantom.
Eq.\  (\ref{cons2tot}) can then be integrated, giving
\begin{equation}
\label{delta0}
\Delta=\Delta_0 a^{-3(1+w_+ +q_-)} +\Delta_{c} a^{-3(1+w_+)} + \Delta_* \,
\end{equation}
where $\Delta_0$ is an integration constant,  $\Delta_c\propto \rho_c$ and $\Delta_* \propto q_0$. This case is of limited interest, because in general the two components  will have a negative energy density in the past or in the future.

Assuming now $w_-\not =0$,
we  obtain from  (\ref{constot}-\ref{cons2tot})  a single equation for the total energy density 
\begin{eqnarray}
\label{rhodiff}
&\rho_{T}'' \, -{\rm tr}({\bf J})\,  \rho_{T}'+\det({\bf J})\, \rho_{T}=9w_{-}q_{0},
\end{eqnarray}
with ${\rm tr}({\bf J})$ and $\det({\bf J})$ respectively given by Eqs.\ (\ref{trJ}) and (\ref{detJ}). 
It is noticeable that - when $\det({\bf J})\not=0$ - this equation admits a constant particular solution corresponding to the constant source term, given by the fixed point (\ref{effL}). Eq.~(\ref{effL}) with  $\det({\bf J})\not=0$  is exactly this constant particular solution, which is the effective cosmological constant term and it vanishes if $q_{0}=0$.

When the eigenvalues (\ref{eigenvals}) of the characteristic polynomial  of Eq.\ (\ref{rhodiff}) are distinct, the general solution can be written as
\begin{eqnarray}
\label{rhotot}
\rho_{T}=\rho_{T+}a^{-3(1+\beta_{+})}+\rho_{T-}a^{-3(1+\beta_{-})}+\rho_{\Lambda}.
\end{eqnarray}
When the eigenvalues are equal, $\lambda_+=\lambda_-=\lambda_0$, we have  the special {\bf Case 1b} of the previous Section  and, denoting $T= {\rm tr}({\bf J})$, we obtain:
\begin{equation}
\label{improper}
\rho_{T}=[\rho_{T1} +\rho_{T2} \ln(a)]\  a^{\frac{T}{2} } + \rho_\Lambda\,.
\end{equation}
Given that this evolution law is similar to the affine one, Eq.\ (\ref{affine}), except for the $\ln(a)$ correction term, and that the fixed point of this {\bf Case 1b} is an improper node, we can refer to this as an improper affine evolution.
Here $\rho_{T+}$, $\rho_{T-}$ and $\rho_{T1}$, $\rho_{T2}$  are  integration constants related to the present values of the energy densities $\rho_A$ and $\rho_B$ of the two dark components;  they are not independent in a flat universe, but related by imposing that the  total energy density has the critical value at present. The effective cosmological constant $\rho_\Lambda$ is given by (\ref{effL}).
In both cases (\ref{rhotot}) and (\ref{improper}) the expansion of the universe is governed by this total energy density that enters the Friedmann equations (\ref{einstein1}-\ref{ray}).
The parameters  $\beta_{\pm}$ are related to the eigenvalues $\lambda_\pm$ by $\lambda_\pm= -3(1+\beta_\mp)$. That is, $\beta_\pm=\beta_{0}\pm\sqrt{D}/3$, with $\beta_{0}=w_{+}+q_{-}/2=-{\rm tr}({\bf J})/6-1$ and  $D$ given by Eq.\ (\ref{D}). 
The case (\ref{improper}) arises from $D=0$, with $\lambda_0=-3(1+\beta_0)={\rm tr}({\bf J})/2$.

It is interesting to note the role of the parameters  $\beta_{\pm}$  in (\ref{rhotot}), assuming they are real. For a barotropic perfect fluid with constant EoS parameter $w$ one has $\rho\propto a^{-3(1+w)}$. Hence, 
$\beta_{\pm}$ simply represent two constant effective EoS parameters that drive the evolution of the total energy density (\ref{rhotot}).  In other words,   the simple coupling (\ref{Coup}-\ref{coupling}) of the two dark components with constant $w_A$ and $w_B$ in general produces a total energy density (\ref{rhotot}) equivalent to that of two uncoupled fluids with constant EoS parameters $\beta_{\pm}$, plus the effective cosmological constant $\rho_\Lambda$ \footnote{Mathematically, this is easily understood in terms of the dynamical system formalism of the previous Section: because the system (\ref{Xprime}) is linear,  it can be globally (in phase space) written in Jordan normal form \cite{KArrowsmith:1992}. When the eigenvalues (\ref{eigenvals})  $\lambda_\pm$ are real and distinct this Jordan form is diagonal, i.e.\  the dynamical system separates  into two normal modes, physically corresponding to two new effective uncoupled fluids   with constant EoS parameters $\beta_{\pm}$. }.

Apart from the asymptotic  effective cosmological constant $\rho_\Lambda$ arising when $q_0\not=0$ and $w_-\not=0$ ({\bf Case 1} of the previous Section), it is also possible to have asymptotic effective cosmological constants when  $\det({\bf J})=0$ ({\bf Case 2}  of the previous Section) if $q_0=0$, corresponding to the vanishing of one of the eigenvalues (\ref{eigenvals}), i.e. $\lambda_{-/+}=0$ and to $\beta_{+/-}=-1$. There are three cases,  given in Table \ref{detJ0}, each with  the corresponding  non vanishing eigenvalue. In this cases $\rho_T$, $\Delta$, $\rho_A$ and $\rho_B$ have the affine behavior (\ref{affine}).  Hence the acceleration is guaranteed by the component of the total energy density for which an effective EoS parameter $\beta_{+/-}$ assumes value $-1$. These fixed points seem less interesting however, because even when the non-zero eigenvalue is negative they are not general attractor of the dynamics (since the other eigenvalue is null): there is a different asymptotic non zero value of $\rho_T$ for each possible initial condition.


We have already commented above on {\bf Case 1b}, giving (\ref{improper}). We now consider the two other sub-cases of {\bf Case 1} of the previous Section from the point of view of the solution (\ref{rhotot}). In both cases $\Delta$, $\rho_A$ and $\rho_B$ evolve with the scale factor as $\rho_T$ in  (\ref{rhotot}), i.e.\ they are all linear combinations of the two normal modes of the system, $a^{-3(1+\beta_{+})}$ and $a^{-3(1+\beta_{-})}$, plus a constant term.

\noindent
{\bf Case 1a}. When both eigenvalues are real and distinct  $\rho_T$  scales as it would if we were considering two decoupled components with EoS parameters $\beta_+$ and $\beta_-$, plus a cosmological constant. If $\det({\bf J})<0$ the fixed point is a saddle, i.e.\ unstable,  with $\lambda_+>0$  and correspondingly $\beta_-<-1$. Consequently  $\rho_T$  scales as  a mixture of  a standard component and a phantom component. When $\det({\bf J})>0$ either both $\beta_{+/-}>-1$, or vice versa, with corresponding standard/phantom behaviour.

\noindent
{\bf Case 1c}. If $\lambda_\pm$ are complex conjugates, then $\beta_{\pm}=\beta_{0}\pm i\sqrt{|D|}/3$ and  the total energy density can be written as
\begin{eqnarray}
\rho_{T}& = & a^{-3(1+\beta_{0})}\left\{(\rho_{T-}-\rho_{T+})\sin{\left[\sqrt{|D|}\ln(a)\right]}\right. \nonumber \\
\label{complex}
&  & \left. +(\rho_{T-}+\rho_{T+})\cos{\left[\sqrt{|D|}\ln(a)\right]}\right\} +\rho_\Lambda
\end{eqnarray}
containing an oscillating function that modulates the power law scaling. 

We will  now investigate the dynamics  of the density parameters. In Sec.~\ref{Markov} we will present a MCMC analysis with SNe data.

\section{Analysis of specific  couplings}
\label{critpoints}

\subsection{Dynamics of density parameters}
\label{dparam}
Introducing an interaction between two fluids can lead to interesting solutions for the energy densities, like attractor points in the phase space where the contributions of the two fluids to the total energy density are constants.  In these points  the value of the normalised energy densities depends only on the parameters of the model and, since they are attractors, they are reached from a wide range of initial conditions,  thereby alleviating the coincidence problem. These are usually called ``scaling solutions'' \cite{Wands:1993,Copeland:1998p1287} and are characterized by constant fractions of the energy density parameters, namely $\Omega_{A,B}=\rho_{A,B}/(3H^{2})$ (in units $8\pi G=1$, $c=1$). 

In order to analyse the dynamics of the system, let us define the new variables:
\begin{eqnarray}
\label{xy}
x=\frac{\rho_{A}}{3H^{2}};\qquad y=\frac{\rho_{B}}{3H^{2}}; \qquad z=\frac{\rho_{\Lambda}}{3H^{2}},
\end{eqnarray}
where together with the coupled fluids we also include radiation to include the era when it's the dominating component, when initial conditions are usually set. 
Note that $x=\Omega_{A}$, $y=\Omega_{B}$ and   $\Omega_\gamma$  are constrained by $x+y+\Omega_{\gamma}=1$; $z$ is the energy density parameter of the total effective cosmological constant, and we neglect the baryons contribution, which is always subdominant. The system (\ref{cons}-\ref{cons2}) then becomes
\begin{eqnarray}
\label{xycons}
x'&=&-x\Big[3(1+w_{+}-w_{-})+2\frac{H'}{H}\Big]\\
\nonumber&+& \frac{3}{2}\Big[(q_{+}-q_{-})x+(q_{+}+q_{-})y+\frac{\det({\bf J})}{9w_{-}}z\Big]\\
\label{xycons2}
y'&=&-y\Big[3(1+w_{+}+w_{-})+2\frac{H'}{H}\Big]\\
\nonumber&-& 3\Big[(q_{+}-q_{-})x+(q_{+}+q_{-})y+\frac{\det({\bf J})}{9w_{-}}z\Big]\\
\label{xycons3}
z'&=&-2z\frac{H'}{H},
\end{eqnarray}
where 
\begin{eqnarray}
\label{eq:Raychaudhuri}\nonumber
\frac{H'}{H} & = & -1-\frac{1}{2} \left[x(1+3(w_{+}-w_{-})) \right. \\
&  & \left. +y(1+3(w_{+}+w_{-}))+2(1-x-y) \right] 
\end{eqnarray}
is a rewriting of the Raychaudhuri equation (\ref{ray}) for the Hubble expansion scalar.

The fixed points, namely the points satisfying $x'=y'=z'=0$,  are presented in Table \ref{table1}, labeled by capital letters,  together with  the corresponding eigenvalues. To the best of our knowlegde, this is the first complete analysis of the dynamics of a three components cosmological system where two of the barotropic fluids are coupled via a general linear coupling function of the form (\ref{coupling}).  The effective EoS parameters at each of the fixed points $w_{eff}=p_{tot}/\rho_{tot}$ is also listed, where $\rho_{tot}=\rho_{A}+\rho_{B}+\rho_{\gamma}$ and therefore $w_{eff}=(w_{+}-w_{-})x+(w_{+}+w_{-})y+\Omega_{\gamma}/3$.


\begin{table*}[htdp]
\begin{center}\begin{tabular}{|c|c|c|c|c|c|c|c|}
\hline  Points & $x$ & $y$ & $z$ & $w_{eff}$ & $\lambda_{x}$ & $\lambda_{y}$ & $\lambda_{z}$ \\
\hline A & $0$ & $0$ & $0$ & $\frac{1}{3}$ & $4$ & $1-3\beta_{+}$ & $1-3\beta_{-}$ \\ 
\hline B & $-\frac{q_{-}-2w_{-}+2\sqrt{D}/3}{4w_{-}}$ & $\frac{(q_{-}-2w_{-}+2\sqrt{D}/3)(q_{+}+2w_{-}+2\sqrt{D}/3)}{4w_{-}(q_{+}+q_{-})}$ & $0$ & $\beta_{+}$  & $3(1+\beta_{+})$ & $-\frac{1-3(\beta_{+}-2\beta_{-}+q_{-})+\sqrt{F_{+}}}{2}$ & $-\frac{1-3(\beta_{+}-2\beta_{-}+q_{-})-\sqrt{F_{+}}}{2}$\\ 
\hline C & $-\frac{q_{-}-2w_{-}-2\sqrt{D}/3}{4w_{-}}$ & $\frac{(q_{-}-2w_{-}-2\sqrt{D}/3)(q_{+}+2w_{-}-2\sqrt{D}/3)}{4w_{-}(q_{+}+q_{-})}$ & $0$ & $\beta_{-}$ & $3(1+\beta_{-})$ & $-\frac{1-3(\beta_{-}-2\beta_{+}+q_{-})+\sqrt{F_{-}}}{2}$ & $-\frac{1-3(\beta_{-}-2\beta_{+}+q_{-})-\sqrt{F_{-}}}{2}$ \\ 
\hline  D & $\frac{1+w_{-}+w_{+}}{2w_{-}}$ & $-\frac{1-w_{-}+w_{+}}{2w_{-}}$ & $1$ & $-1$ & $-4$ & $-3(1+\beta_{+})$ & $-3(1+\beta_{-})$ \\ 
 \hline
\end{tabular} 
\caption{ \small \small Fixed points of system (\ref{xycons}-\ref{xycons3}),  the corresponding effective EoS and eigenvalues, where $F_{\pm}=9 q_{-}^{2}/2 - 3 q_{-}(1\pm \sqrt{D} - 3w_{+}) + 9q_{+}w_{-} + 9w_{-}^{2} - (1\pm 2\sqrt{D} - 3w_{+})(-1 + 3w_{+})$.\label{table1} }
\end{center}
\end{table*}
%
All the fixed points shown in Table~\ref{table1} exist for $w_{-}\neq0$, when the EoS parameters of the two fluids are the different. As aforementioned, the only physically reasonable fixed point for system (\ref{constot}-\ref{cons2tot}) corresponding to $w_{-}=0$ is {\bf Case 2a}, where $\det({\bf J})=0$.  
  
\begin{figure}[ht!]
  \centering
    \includegraphics[width=7.truecm]{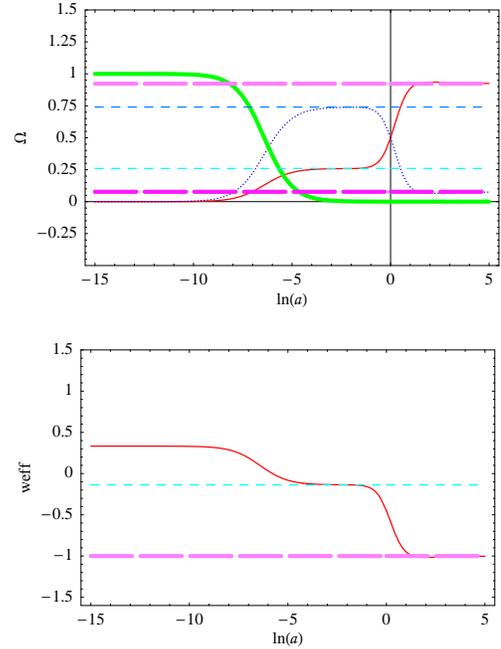}
    \caption{ \small \small Upper panel: evolutions for the energy density parameters  for a model with $q_{-}=0$ and $q_{+}=-0.5$; for comparison, the dashed lines are the values of $x$ and $y$ at  the fixed points B (thin short-dashed lines) and D (thick long-dashed lines). For this model the parameters are: $\Omega_{0A}=\Omega_{\Lambda}=0.5$,  $w_{A}=-1.1$, $w_{B}=0.2$.  Lower panel: effective EoS for the same model; for comparison, we plot  the  EoS parameter of the fixed point B, $\beta_{+}=-0.14$.}
    \label{fig:omegasbis}
\end{figure}

\begin{figure}[ht!]
  \centering
    \includegraphics[width=7.truecm]{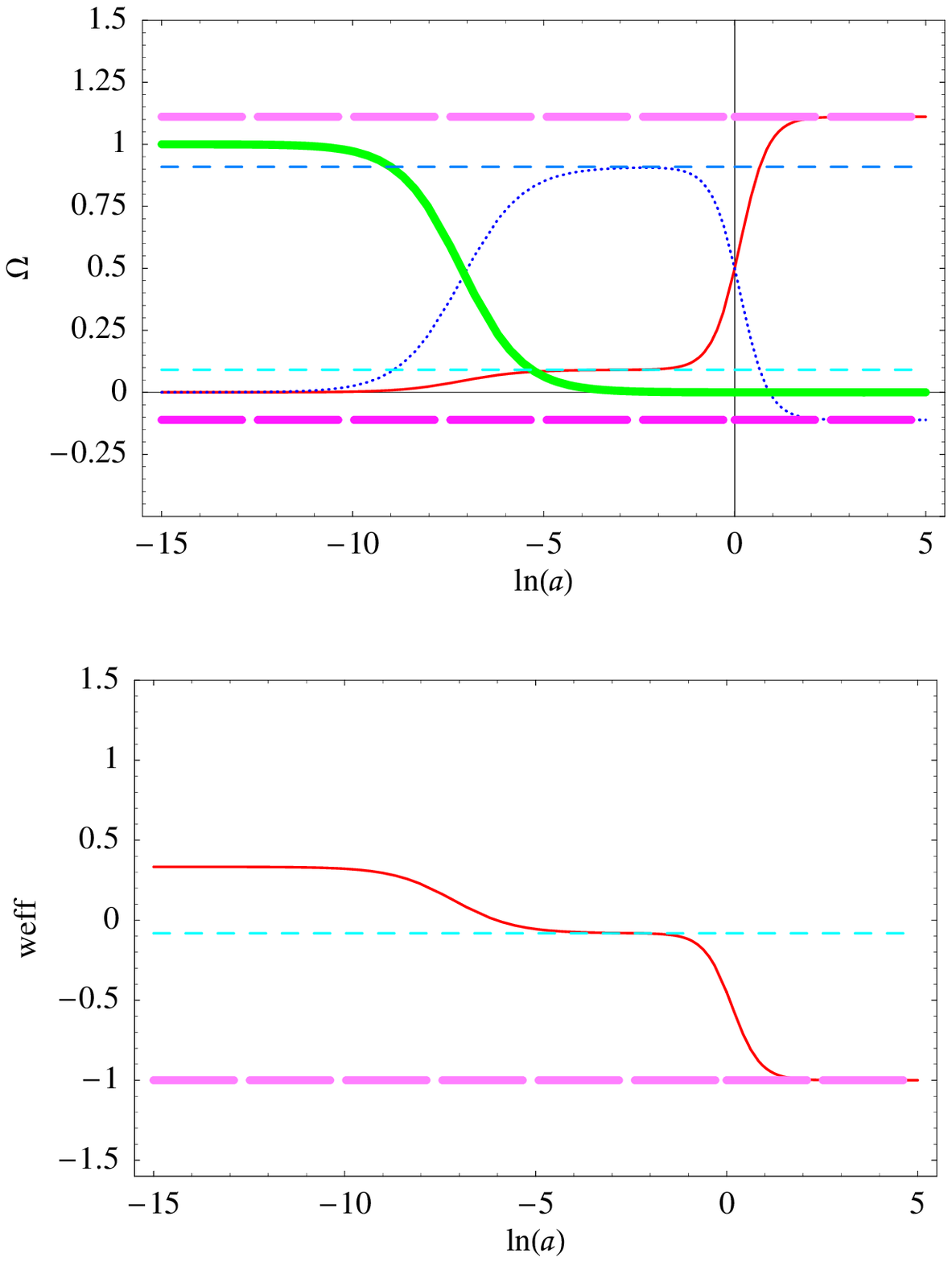}
    \caption{ \small \small Upper panel: evolutions for the energy density parameters  for a model with $q_{+}=0$ and $q_{-}=-0.18$; for comparison, the dashed lines are the values of $x$ and $y$ at the fixed point B (thin short-dashed lines) and D (thick long-dashed lines). For this model the parameters are: $\Omega_{0A}=0.5$, $\Omega_{\Lambda}=0$,  $w_{A}=-0.9$, $w_{B}=0.$ (dust). The  EoS parameters at  B are $\beta_{+}=-0.08$ and $\beta_{-}=-1$. Lower panel: effective EoS for the same model.  }
    \label{fig:omegastris}
\end{figure}

The fixed points A corresponds to the radiation dominated era, while B, C and D represent epochs that are dominated by the two fluids. In particular, at the fixed point D  the constant energy densities of $x$ and $y$ ($\rho_{A\Lambda}$ and $\rho_{B\Lambda}$) cause the accelerated expansion with $w_{eff}=-1$. From the expression for $x$ and $y$ at this latter fixed point it is easy to see that $y=-x(1+w_{A})/(1+w_{B})$, which is exactly the proportionality that holds for {\bf Case 1} and {\bf Case 2c}, as discussed at the end of Sec.~\ref{stab}: this point is characterized by the final domination of an effective cosmological constant, either driven by $q_{0}$ ({\bf Case 1}, see Fig.~\ref{fig:omegasbis}) or not ({\bf Case 2c}, see Fig.~\ref{fig:omegastris}). In the first case whenever $|\beta_{\pm}|<1$ D is always an attractor, while in the second case  it is not because one of the eigenvalues is null. Notice that its existence is completely independent on $q_{+}$ and $q_{-}$. Whenever the system settles into the fixed points B or C the role of $\beta_{+}$ and $\beta_{-}$ is exactly that of  effective EoS parameters (see Table~\ref{table1}) which allows for phantom line crossing at late time (see Fig.~\ref{fig:omegas}), i.e.\ line for which the effective total  EoS parameter is $w_{eff}=-1$. 

In the following we will examine in more detail three special classes of the coupling function and in Sec.~\ref{Markov} we will make a first comparison of the models to the data using MCMC  applied to type Ia SNe distance modulus.

\subsection*{I. $q_{+}=q_{-}$}
\label{single} 
Imposing $q_{+}=q_{-}$ is equivalent to choosing $q_{A}=0$ and $q_{B}=q$; therefore among the range of possible couplings represented by (\ref{coupling}) we are restricting to the class of models where $Q/H$ is proportional solely to the energy density of one fluid (in our case e.g.\ $\rho_{B}$), and it reads
\begin{equation}
\label{coupl1}
\frac{Q}{H}=\frac{3}{2}(q\rho_{B}+q_{0}).
\end{equation}
This assumption also includes models with $q_{+}=-q_{-}$ since the coefficients $q_{A}$ and $q_{B}$ can be either negative of positive. In this case the dynamics is the same as for $q_{+}=q_{-}$, the roles of $x$ and $y$ being simply interchanged. We will refer to this subclass of models as model I.

In this model $\sqrt{D}$ is automatically real, since $D=9(q/2+w_{B}-w_{A})^{2}/4$; as a consequence the scaling function (\ref{rhotot}) always drives a power law expansion, with $\beta_{+}=q/2+w_{B}$ and $\beta_{-}=w_{A}$ if $\beta_{0}>0$ (i.e.\ $(w_{B}+w_{A}+q/2)>0$),  vice versa if $\beta_{0}<0$.
Hence the total fluid ends up as if it was  made up of: {\it i)} a component scaling as the original  fluid  $\rho_{A}$ with no coupling, {\it ii)} a second component characterized by a new EoS parameter and {\it iii)} an effective cosmological constant term $\rho_{\Lambda}$. Moreover a pure affine behaviour (\ref{affine}), or its improper modification (\ref{improper}), is  obtained in three cases: {\it i}) $q=-2(w_{B}-w_{A})$, which gives (\ref{improper}); {\it ii})  $q=-2(1+w_{B})$, that corresponds to $\beta_{+/-}=-1$ (even for $\rho_{\Lambda}=0$, i.e.\ $q_{0}=0$, an effective cosmological constant is generated); {\it iii}) $w_{A}=-1$, where one of the two fluids is {\it ab initio} a constant term.  
Notice however that generally, because $\beta_{+}=\beta_{-}+2\sqrt{D}/3$, models with $\beta_{+}=-1$ and $\beta_{-}>-1$ are not feasible.  In particular, the $\Lambda$CDM evolution is exactly recovered in case {\it ii})  for $w_{A}=0$, that is if one of the fluids {\it is } dust; in case {\it iii})  for $q_{B}=-2w_{B}$.

\begin{figure}[ht!]
  \centering
    \includegraphics[width=7.truecm]{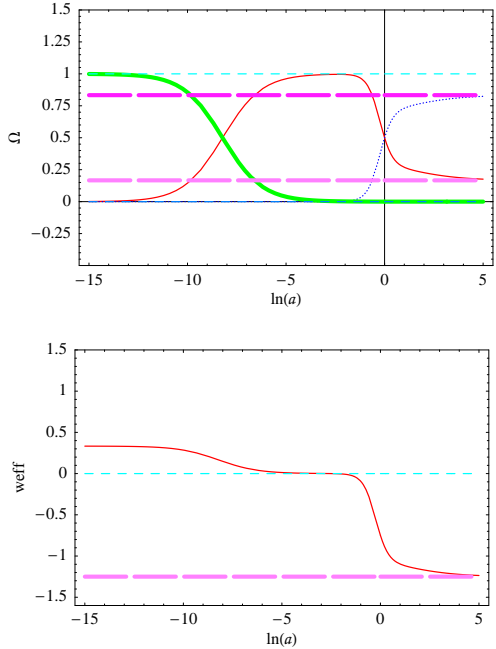}
    \caption{ \small \small Upper panel: evolutions of the energy density parameters $\Omega_{A}$ (thin solid line), $\Omega_{B}$ (dotted line) and $\Omega_{\gamma}$ (thick solid line) for a model with $q_{+}=q_{-}=0.25$; for comparison, the dashed lines are the values of $x$ and $y$ at  the fixed points B (thin short-dashed lines) and C (thick long-dashed lines). For this model the parameters are: $\Omega_{0A}=\Omega_{\Lambda}=0.5$, $w_{A}=0$, $w_{B}=-1.5$,  $\beta_{+}=0$ and $\beta_{-}=-1.25$.  Lower panel: the total effective EoS parameter for the same model : $w_{eff}$ evolves from the value $1/3$ in the radiation dominated era, approaches the value $~0$ in the matter dominated era and then asymptotically evolves toward a constant phantom value, in this case  $\beta_{-}=-1.25$. }
    \label{fig:omegas}
\end{figure}

The fixed point D is characterized by the domination of the constant part of the total energy density $\rho_{\Lambda}$; along it, the values of $x$ and $y$ are both positive only if either $w_{A}$ or $w_{B}$ have phantom values, i.e.\ $w_{A}<-1$ or $w_{B}<-1$. This statement holds true also for models II and III. However, if $w_{A}<-1$ D is no longer an attractor, as $\lambda_{y}=-3(1+w_{A})$ is greater than zero. On the other hand $w_{B}<-1$ requires $q>-2(1+w_{B})$ to let the fixed point be an attractor: in this case $q$ is positive. A strong and positive $q$ corresponds to a transfer of energy from $\rho_{A}$ to the other fluid with $w_B<-1$. Therefore in order to fall at late time into the cosmological constant dominated era a fluid with a phantom EoS parameter $w_B$ must absorb energy from the other non-phantom fluid. It is worth stressing that the effective cosmological constant, i.e.\ $q_{0}$, is somewhat redundant whenever the fixed point D is not an attractor (see Fig.~\ref{fig:omegas}). In Fig.~(\ref{fig:omegas}) an  example of this dynamics of the background is shown; the effective cosmological constant is not noticeable, since, after the evolution on the saddle point B, the system is trapped in the attractor point C.


%
\subsection*{II. $q_{-}=0$}
\label{total} 
If $q_{-}=0$ the resulting coupling function $Q/H$ is linearly dependent on the sum of the energy densities of the two fluids, approximately equivalent to the total energy density (these models have been examined for example in \cite{Olivares:2006} and \cite{Abdalla:2007}) and is as follows
\begin{equation}
\label{coupl2}
\frac{Q}{H}=\frac{3}{2}(q\rho_{T}+q_{0}).
\end{equation}
With this assumption $q_{A}=q_{B}=q_{+}=q$ and $\beta_{\pm}=w_{+}\pm\sqrt{D}/3$ where $D=9(w_{B}-w_{A})(2q+w_{B}-w_{A})/4$. If $q$ is positive these effective EoS are real for $w_{B}> w_{A}$ or $w_{B}\le w_{A} -2q$, while if $q$ is negative the same relations hold but with opposite inequality signs. We will label this model II.

In this model the affine evolution is recovered for $w_{B}=(qw_{A}+2w_{A}+2)/(q-2w_{A}-2)$, corresponding to  $\beta_{-}=-1$. In this case, which is indeed {\bf Case 2c} of Section  \ref{coupledfluids}, an effective cosmological constant arises even for $\rho_{\Lambda}=0$. Again, because $\beta_{+}=\beta_{-}+2\sqrt{D}$, models with $\beta_{+}=-1$ and $\beta_{-}>-1$ are not feasible. From a cosmological point of view this means that a matter-like evolution cannot be generated together with a cosmological constant. The $\Lambda$CDM limit is achieved if$w_{A}=(-1+q\pm\sqrt{q^{2}+1})/2$ and $w_{B}=-1-w_{A}$. The evolution of the energy densities for a special choice of the parameters is illustrated in Fig.~\ref{fig:omegasbis}: the effective cosmological constant (\ref{effL}) arises at late time, driving the acceleration, and $\rho_{\Lambda}$ is caused by a non-zero $q_{0}$ ($\Omega_{\Lambda}\neq 0$).

\subsection*{III. $q_{+}=0$}
\label{difference} 
The subgroup of models with $q_{+}=0$ ( from now on model III) includes the couplings that are proportional to the difference of the energy densities $\Delta$ (for example recently analysed in \cite{Chimento:2007bis}). With this assumption $q_{-}=q_{B}=-q_{A}=q$ and the discriminant $D=9(q^{2}+(w_{B}-w_{A})^{2})/4$ is always positive, so that oscillating solutions (\ref{complex}) are never permitted. The coupling function reads
\begin{equation}
\label{coupl3}
\frac{Q}{H}=\frac{3}{2}(q\Delta+q_{0}).
\end{equation}
As before,  the affine expansion (\ref{affine}) may only be generated if one of the two effective EoS parameters assumes the value of the EoS of a cosmological constant, that is either $\beta_{+}=-1$ or $\beta_{-}=-1$. In particular if $\beta_{+}=-1$, $\beta_{-}=-1-2\sqrt{D}/3$ is always  phantom. In this case none of the terms in Eq.~(\ref{rhotot}) can play the role of matter. On the other hand if $\beta_{-}=-1$ (corresponding to $w_{A}=(-1-q\pm\sqrt{1-q^{2}})/2$), $\beta_{+}=-1+2\sqrt{D}/3$ is always grater than $-1$, i.e.\ always standard. An example of this dynamics is shown in Fig.~\ref{fig:omegastris}, where the effective cosmological constant (\ref{effL}) arises at late time with no need of $q_{0}$, driving the acceleration ({\bf Case 2c}). Then typically for $w_{B}=-q-w_{A}-1$ we have that $\beta_{+}=0$ and  the $\Lambda$CDM model is recovered.

\section{Markov chains with supernovae}
\label{Markov}
\subsection{Methods}
\label{methods}
Given the large number of parameters, the task of finding the minimum $\chi^{2}$ and mapping its distribution in the entire parameter space can be computationally expensive. To this end we adopt a MCMC. In this work we only want to test our models as a description of the homogenous isotropic background expansion (regardless of perturbations), hence supernov\ae \ are ideal for this purpose.  We use the 192 type Ia SNe distance modulus data set provided in  \cite{Davis:2007}. In particular we want to see wether  supernovae can qualitatively distinguish different kind of couplings, included what we called model I, II and III.

Type Ia SNe light curves allow a determination of an extinction-corrected distance moduli,
\begin{equation}
\mu_0=m-M=5\log \left(d_L/{\rm Mpc}\right)+25
\label{distance_modulus}
\end{equation}
where $d_L=(L /4\pi F)^{1/2}=(1+z)\int_0^z dz' / H(z')$ is the luminosity distance. We compare our theoretical predictions to the values of $\mu_0$ with $H^{2}=8\pi G/3(\rho_{A}+\rho_{B}+\rho_{\gamma}+\rho_{b})$, where we account also for the baryon energy density $\rho_{b}$. We fix the value of the  dimensionless Hubble constant to be $h=0.72$ \cite{HSTKeyP01} and the baryon energy density at present $\Omega_{b}h^{2}=0.02229$  according to  \cite{Spergel:astro-ph/0603449}. The smaller is the EoS parameter of a single fluid the later can be the domination era for this fluid. Hence, counting the role of $\beta_{\pm}$ as effective EoS parameters, whenever $\beta_{\pm}>0$ a baryonic era might emerge at recent time. The absolute distance modulus $M$ is intrinsically affected by uncertainty; therefore we treat it as a nuisance parameter and marginalize over it.

The parameters that are representative of the models are $\{\Omega_{0A},\Omega_{\Lambda},q_{A},q_{B},w_{A},w_{B}\}$, or otherwise $\{\Omega_{0A},\Omega_{\Lambda},q_{+},q_{-},w_{+},w_{-}\}$ and, as functions of these, the two effective  EoS introduced in Eq.~(\ref{rhotot}): $\beta_{+}$ and $\beta_{-}$.  For the ensuing analysis it is worth reminding our classification of models: I) model with a coupling function proportional to only one of the two energy densities; II) model with a coupling function proportional to the sum of the energy densities; III) model with a coupling function proportional to the difference of the energy densities.

We shall now focus our analysis on the case  $w_{B}=0$, i.e.\  $\rho_B$  would represent standard CDM if it wasn't for the coupling with the DE component.

\subsection{Results:  CDM - DE coupled models}
\label{results}
The first result we obtain is that $\Omega_{\Lambda}$ is completely unconstrained, independently on which model we consider. This means that SNe are not sensitive to the  constant term of the coupling.  As we have seen  in Sec.~\ref{coupledfluids} the dynamics of the system can easily generate the acceleration settling on fixed points D, where $w_{eff}=-1$, even for $\Omega_{\Lambda}=0$ (see Fig.~\ref{fig:omegastris}), or B and C, where the total energy density can also exhibit phantom evolutions. 

\begin{figure*}[!ht]
  \centering
    \includegraphics[width=17.truecm]{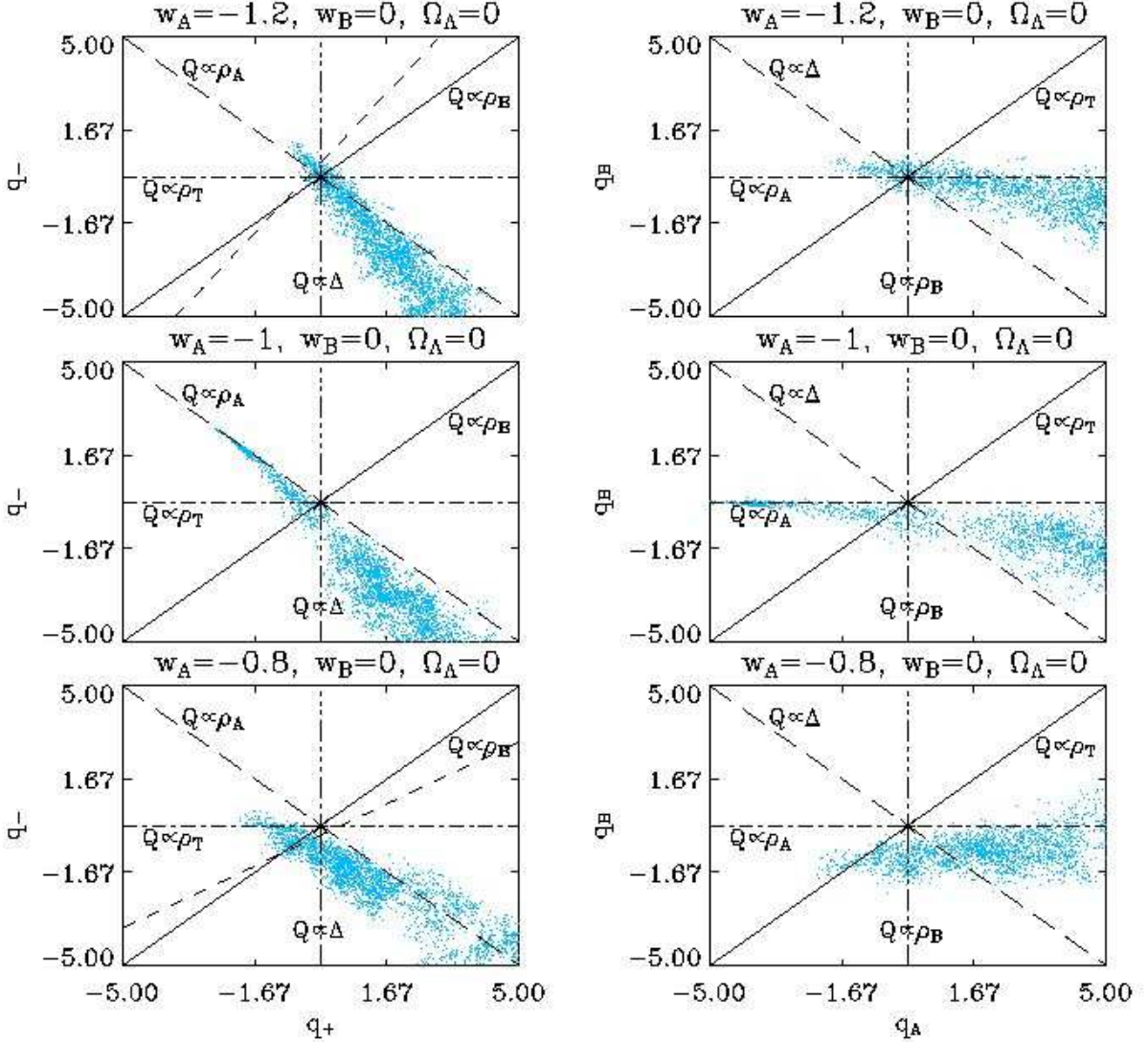}
    \caption{ \small \small Coupling diagrams with two-dimensional likelihood for models with $\Omega_{\Lambda}=0$. Apart from the short-dashed line that represents an affine evolution with $\beta_{+/-}=-1$, all the other lines are labeled with the corresponding type of coupling function (e.g.\ the solid line on the left side diagrams represents a coupling function $Q\propto\rho_{B}$ (model I), while on the right side diagrams it represents $Q\propto \rho_{T}$ (model II)). The energy density parameter at present is fixed at its best fit value, respectively  $\Omega_{0A}=0.63, 0.65, 0.76$. }
    \label{fig:res1}
\end{figure*}

\begin{figure*}[!ht]
  \centering
    \includegraphics[width=17.truecm]{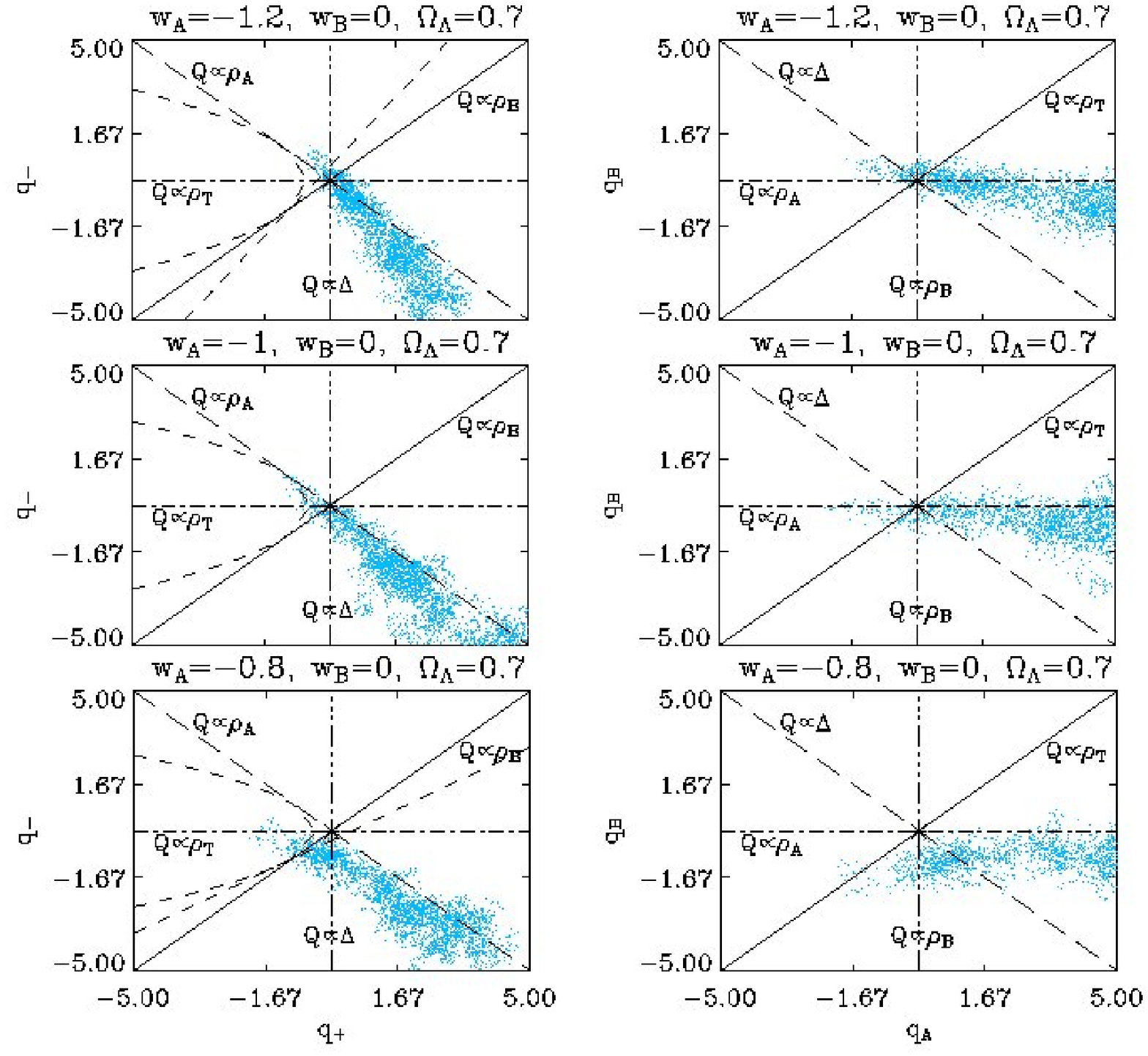}
    \caption{ \small \small  Coupling diagrams with two-dimensional likelihood for models with $\Omega_{\Lambda}=0.7$.  All the lines are labeled with the corresponding type of coupling function (e.g.\ the solid line on the left side diagrams represents a coupling function $Q\propto\rho_{B}$ (model I), while on the right side diagrams it represents $Q\propto \rho_{T}$ (model II)). The short-dashed line represents affine evolution with $\beta_{+/-}=-1$ and the short-dashed curve represents affine evolution with $\beta_{+}=\beta_{-}$ . The energy density parameter at present is fixed at its best fit value, respectively  $\Omega_{0A}=0.63, 0.65, 0.76$. }
    \label{fig:res2}
\end{figure*}

In Fig.~\ref{fig:res1} and \ref{fig:res2} we present MCMC chains in a two-dimensional diagram $[q_{+},q_{-}]$ ($[q_{A},q_{B}]$ on the right hand side). As said above, we consider a model where one of the two fluid represents a CDM component, i.e.\ $w_{B}=0$, a reasonable assumption considered all the other cosmological probes pointing towards the existence of a form of cold dark matter (see e.g.\ \cite{Khalil:2002}), and we let $w_{A}$ assume three different values that characterize $\rho_{A}$ as a DE component (phantom-like behaviour is shown in the top panels, cosmological constant-like in the second row panels and non-phantom model in the bottom panels).

Note that the case where $w_{A}=0$ and $\rho_{B}$ is DE can be easily derived from the previous one, corresponding in the diagram to a reflection with respect to the line $q_{B}=-q_{A}$. In fact, interchanging the two EoS and swapping the roles of the two energy densities, and applying the transformation ($q_{A}\rightarrow -q_{B}$, $q_{B}\rightarrow -q_{A}$, i.e.\ $q_{+}\rightarrow -q_{+}$, $q_{-}\rightarrow q_{-}$), one recovers the aforementioned model.

In addition, the straight lines corresponding to models I, II and III are drawn, and diagrams of Fig.~\ref{fig:res2} are derived from the same choice of parameters as in Fig.~\ref{fig:res1} except for $\Omega_{\Lambda}\neq 0$ (it is by eye easily verifiable that there is no dependence on $\Omega_{\Lambda}$). Finally, the short-dashed curves represent the improper affine evolution (\ref{improper}),  while the short-dashed straight line represents affine models (\ref{affine}) with $\beta_{+/-}=-1$ ({\bf Case 2c}).

As a first step we derived the unidimensional likelihood for $\{\Omega_{0A},q_{+},q_{-}\}$. The best fit of the energy density parameter for the three class of models presented in Fig.~\ref{fig:res1} and \ref{fig:res2} is respectively $\Omega_{0A}=0.63,0.65,0.76$ with an error of $2\sigma=0.1$; this best fit does not change including $\Omega_{\Lambda}$. In the diagrams $\Omega_{0A}$ is therefore fixed to these best fit values. It is worth stressing that here we are not just analysing the typical models considered in literature (namely I, II and III) but the results incorporate {\it all} the possible linear couplings, and, we might say, all the possible expansions at recent times of a generic coupling function $Q$ (Eq.~\ref{Coup}). Hence we are not interested in deriving constraints on single parameters, a route that might be hard to follow with SN Ia in view the high number of parameters and their degeneracies. We instead want to see what kind of linear couplings are preferred by the data and provide a qualitative way to distinguish the type and the direction of the interaction.

The first noticeable thing in the $[q_{A},q_{B}]$ diagram is that the points lie almost on a horizontal branch of the diagram, close to the line representing model I, in particular with $Q\propto \rho_{A}$. So if we allow the interaction term to be strong and move out of the weak coupling regime (i.e.\ \tbf{$|q_{A,B}|>1$}),  the most ``frequent'' linear coupling function emerging from the chains is the one proportional to the DE density ($\rho_{A}$). In addition, strong couplings are favoured for positive value of $q_{A}$ (see Fig.~\ref{fig:res1} and \ref{fig:res2}): the energy is transferred from dark matter to DE. Increasing the value of $w_{A}$, that is moving from phantom-like values towards  quintessence-like ones (going downwards in the right hand side column of fig.~(\ref{fig:res1})), this horizontal branch tends to negative values of $q_{B}$. Models with a  phantom $w_A$  show an increasing energy density with the scale factor, $a$, while for DE model characterized by $w_A>-1$ the energy density is diluted with the universe expansion: this second kind of models requires a lower transfer of energy from CDM to DE. Apart from a small spot in the origin of the axis (weak couplings), the coupling or type  II does not seem to be favored by SN data, the effect increasing with higher values of  $w_{A}$, i.e.\ for non-phantom values. Another evidence that arises from diagrams Fig.~\ref{fig:res1} and \ref{fig:res2} is that for non-phantom values of $w_A$ the uncoupled  case (namely $[q_{A},q_{B}]=[0,0]$) falls almost outside the border of the likelihood. 

Since today $\rho_{0B}\simeq \rho_{0A}$ we can say that the sign of the coupling function ($Q \simeq q_{A}\rho_{0A}+q_{B}\rho_{0B}$) changes along the straight line $q_{B}=-q_{A}$ (long dashed line): above this line the exchange term reverses the energy transfer from CDM to DE (i.e.\  positive $Q$), while below it is the opposite (negative $Q$). Again, the higher is $w_{A}$ the bigger is the number of points that we can find below this line. Therefore for DE components with $w_A<-1$ an exchange of energy from DE to CDM is less probable, independently on the type of linear coupling.  This reflects the fact that an increasing energy density (characteristic of phantom behaviour) favors more and more  absorbing and positive DE couplings at present, while non-phantom values of $w_A$ seem to need a negative exchange term, most of all for weak couplings, to explain supernovae data. It is worth stressing that eventually the likelihood seems to exclude the uncoupled case. 

 The connection between where the points lie in the diagrams, i.e.\ the region favoured by the likelihood, and where the cosmological background evolution is affine is an interesting issue; this directly connects coupled DE models to an effective evolution of the total energy density that  is completely equivalent to a cosmological constant plus a component with constant EoS parameter $\alpha$, Eq.\ (\ref{affine}).
If one looks at the left side diagrams of Fig.~\ref{fig:res2}, a short-dashed curve and a short-dashed straight line are drawn on it. The former corresponds to the improper affine evolution (\ref{improper}), obtained for $q_{-}=\pm\sqrt{-4w_{-}(q_{+}+w_{-})}$.
  Hence the only affine models are those that correspond to  the straight line for witch $\beta_{+/-}=-1$ (for the model with $w_{A}=-1$ this coincides with the line representing model II, the only possibility to recover the affine evolution with no coupling).
For a  DE model with a phantom $w_A$  the affine evolution coexisting with a non-zero $\Omega_{\Lambda}$ is somewhat ruled out and, among the models indicated in the last Section, is more compatible with a coupling function proportional to $\rho_{A}$ (DE, model I) and possibly to $\Delta$ (model III). For DE models with $w_A>-1$  the situation is different: the data seem to favor an affine evolution generated in models with a coupling function proportional to $\rho_{B}$ (matter, model I) and again model III. In addition, for  DE models with standard $w_A$ an improper affine evolution together with a non-vanishing $q_{0}$ ($\Omega_{\Lambda}$) is allowed, in a region where the coupling function shifts towards negative sign, thus representing a transfer of energy from DE to CDM.

\section{Conclusions}
\label{conclusion}
We have analysed the dynamics of two coupled dark components  represented by two  barotropic perfect fluids characterized by  constant EoS parameters $w_A$ and $w_B$. We have assumed a flat, homogeneous and isotropic cosmology  and  a general linear coupling between the two  barotropic perfect fluids. This scale-independent coupling takes a linear form proportional to the single energy densities plus a constant term: any coupling of this type can approximate at late time a more general coupling function. We have studied the stability of the system and shown that an effective cosmological constant can arise both from the constant part  $q_{0}$ of the function $Q$ and from an effective cosmological constant-like EoS. We have also examined the dynamics of  the  energy density parameters, and evaluated the  fixed points and the corresponding eigenvalues,  for the most  general form of linear coupling. We have then restricted the analysis to some specific  linear couplings previously considered in the  literature (model I, II, III). Since we are restricting to the background expansion and we have modeled the coupling function as a late time first order Taylor expansion, a comparison with distance modulus from SN Ia data appeared as our natural step further. We have presented a MCMC analysis for a model with dark matter plus DE  using the data set provided in  \cite{Davis:2007}. Considering two representative specific values of the DE parameter $w_A$, one standard ($w_A>-1$) and the other phantom ($w_A<-1$), we have condensed  our results  in coupling diagrams, where the points arising from the MCMC  chains are drawn together with lines for model I, II and III and for the improper affine (\ref{improper})  affine  (\ref{affine})  evolutions, the latter including  the $\Lambda$CDM model as a subcase.  Couplings proportional to the DE density seem favored, mostly  for strong couplings $|q_A|>1$. The total sign of the exchange term sets the direction of the interaction: models with phantom $w_A$ definitely prefer positive coupling, i.e.\ an energy transfer from dark matter to DE. On the other hand,  models with non-phantom $w_A$ not only allow for negative $Q$, but forces the uncoupled model to fall at the border of the likelihood. For further and stronger constraints  more complementary data are required, like CMB spectra or matter power spectra. These observables necessitate an accurate relativistic perturbation analysis which is nor obvious neither uniquely defined in phenomenological coupled models as those considered here. Moreover, simplified observable that make no use of perturbation analysis, like the CMB shift parameter, can be strongly model-dependent and, although straightforward, should not be used in models where the evolution, even just that of the unperturbed background, detaches significantly from that of the $\Lambda$CDM model. These extended investigations can only be settled with future work. 

\subsection{Acknowledgements} 
We are gratuful to Roy Maartens, Betta Majerotto, Jussi Valiviita and other members of ICG (Portsmouth) for useful discussions. MB work was partly funded by STFC.


\bibliography{Bcoup}
\end{document}